\documentstyle[12pt]{article}
\def\A{{\cal A}}
\def\Op{{\cal O}}
\def\U{{\cal U}}
\def\T{{\cal C}}

\def\H{{\cal H}}
\def\P{{\cal P}}
\def\ot{\otimes}
\def\ra{\rangle}
\def\la{\langle}
\def\tr{\triangleright}

\newcommand{\newsection}{ \setcounter{equation}{0} \section}

\def\appendix#1{
  \addtocounter{section}{1}
  \setcounter{equation}{0}
  \renewcommand{\thesection}{\Alph{section}}
  \section*{Appendix \thesection\protect\indent #1}
  \addcontentsline{toc}{section}{Appendix \thesection\ \ \ #1}
  }
\newcommand{\beq}{\begin{equation}}
\newcommand{\eeq}{\end{equation}}
\newcommand{\bea}{\begin{eqnarray}}
\newcommand{\eea}{\end{eqnarray}}

\newcommand{\zet}{\hbox{\bf Z}}

\newcommand{\hub}{Hubbard model}
\newcommand{\kju}{{K_j^{(+)}}}

\newcommand{\klu}{{K_l^{(+)}}}
\newcommand{\kld}{{K_l^{(-)}}}
\newcommand{\klz}{{K_l^{(z)}}}

\newcommand{\kz}{{K^{(z)}}}
\newcommand{\krz}{{K_r^{(z)}}}
\newcommand{\ku}{{K^{(+)}}}
\newcommand{\kd}{{K^{(-)}}}
\newcommand{\kjz}{{K_j^{(z)}}}
\newcommand{\kiz}{{K_i^{(z)}}}

\newcommand{\aiu}{{a_{i\uparrow}}}
\newcommand{\aid}{{a_{i\downarrow}}}
\newcommand{\aaiu}{{a^{\dagger}_{i\uparrow}}}
\newcommand{\aaid}{{a^{\dagger}_{i\downarrow}}}
\newcommand{\alu}{{a_{l\uparrow}}}
\newcommand{\ald}{{a_{l\downarrow}}}
\newcommand{\aalu}{{a^{\dagger}_{l\uparrow}}}
\newcommand{\aald}{{a^{\dagger}_{l\downarrow}}}

\newcommand{\bu}{{b_{\uparrow}}}
\newcommand{\bd}{{b_{\downarrow}}}

\newcommand{\nld}{{n_{l\downarrow}}}
\newcommand{\nlu}{{n_{l\uparrow}}}

\newcommand{\aju}{{a_{j\uparrow}}}
\newcommand{\ajd}{{a_{j\downarrow}}}
\newcommand{\aaju}{{a^{\dagger}_{j\uparrow}}}
\newcommand{\aajd}{{a^{\dagger}_{j\downarrow}}}

\newcommand{\nid}{{n_{i\downarrow}}}
\newcommand{\njd}{{n_{j\downarrow}}}
\newcommand{\niu}{{n_{i\uparrow}}}
\newcommand{\nju}{{n_{j\uparrow}}}

\newcommand{\ais}{{a_{i\sigma}}}
\newcommand{\bis}{{b_{i\sigma}}}
\newcommand{\aais}{{a^{\dagger}_{i\sigma}}}
\newcommand{\bbis}{{b^{\dagger}_{i\sigma}}}

\newcommand{\bjs}{{b_{j\sigma}}}
\newcommand{\aajs}{{a^{\dagger}_{j\sigma}}}
\newcommand{\bbjs}{{b^{\dagger}_{j\sigma}}}

\newcommand{\nis}{{n_{i\sigma}}}

\newcommand{\ei}{{e^{-i(\vec{\Phi}-\vec{\kappa}) \cdot
\vec p_i-i \vec{\kappa} \cdot \vec p_j+\zeta \vec R_{ij}
   \cdot (\vec x_i-\vec x_j)}}}
\newcommand{\ej}{{e^{-i(\vec{\Phi}-\vec{\kappa}) \cdot
\vec p_j-i \vec{\kappa} \cdot \vec p_i+\zeta \vec R_{ij}
   \cdot (\vec x_i-\vec x_j)}}}
\newcommand{\epi}{{e^{\frac{1}{2} \vec R_{ij} \cdot  \vec{\Phi} \zeta \hbar}}}
\newcommand{\epj}{{e^{-\frac{1}{2} \vec R_{ij} \cdot \vec{\Phi} \zeta \hbar}}}

\newcommand{\pij}{\prod_{r<i}e^{\alpha \krz}
\prod_{r>i, r \not= j}e^{-\alpha^* \krz}}
\newcommand{\pji}{\prod_{r<j,r \not= i}e^{\alpha \krz}
\prod_{r>j}e^{-\alpha^* \krz}}
\newcommand{\eaijm}{e^{-\frac{1}{2}\alpha}}
\newcommand{\eajim}{e^{\frac{1}{2}\alpha^*}}

\newcommand{\hh}{{H_{Hub}}}

\begin{document}
\topmargin 0pt
\oddsidemargin 5mm
\headheight 0pt
\headsep 0pt
\topskip 9mm

\addtolength{\baselineskip}{0.20\baselineskip}
\begin{titlepage}
\begin{center}
September 1995        \hfill       LMU-TPW 95-11\\
\mbox{}            \hfill       cond-mat/9510074\\
\vskip.6in

{\large \bf On Quantum Groups in the Hubbard Model with Phonons}

\vskip.6in

B.~L.~Cerchiai* and P.~Schupp

\vskip.25in

{\sl Sektion Physik der Ludwig-Maximilians-Universit\"at M\"unchen\\
Theoretische Physik --- Lehrstuhl Professor Wess\\
Theresienstra\ss e 37, D-80333 M\"unchen\\
Federal Republic of Germany}
\end{center}
\vskip1in
\begin{abstract}
The correct Hamiltonian for an extended Hubbard model
with quantum group symmetry
as introduced by Montorsi and Rasetti is derived for a
\mbox{$D$-dim}\-en\-sio\-nal lattice. It is shown that the superconducting
SU${}_q$(2)  holds as a true quantum symmetry
only for $D = 1$ and that terms of higher order in the
fermionic operators are needed in addition to phonons.
A discussion of Quantum symmetries in general is given in a
formalism that should be readily accessible to non-Hopf algebraists.
\end{abstract}
\vfill
\noindent \hrule
\vskip.2cm
\noindent {\small *DAAD-fellow \hfill {\it e-mail: }
Bianca, Schupp \ @ \ lswes8.ls-wess.physik.uni-muenchen.de}
\end{titlepage}
\newpage
\setcounter{page}{1}

\newpage

\newsection{Introduction}
The {\hub} is the simplest description of itinerant interacting electron
systems. In this article we will study  generalizations of it
on a $D$-dimensional lattice.
The Hamiltonian of the standard {\hub} is given by~\cite{hubbard}:
\beq
{\hh}=H_{el}^{(non-loc)}+H_{el}^{(loc)}
\eeq
where
\beq
H_{el}^{(non-loc)}= -t \sum_{<i,j>,\sigma} \bbis \bjs,  
\eeq
\beq
H_{el}^{(loc)}=u \sum_i \niu \nid-\mu \sum_{i,\sigma} \nis \label{elloc}.
\eeq
The 1-dimensional model has been solved in~\cite{LW}.
It is well-known that the {\hub} has a $(\mbox{SU}(2)
\times \mbox{SU}(2))/\zet_2$-symmetry
\cite{yangzhang,Yang2}.
This symmetry is the product of two separate $\mbox{SU}(2)$ symmetries:
a magnetic and a
superconductive one.

Montorsi and Rasetti \cite{rasetti} have introduced a very interesting
generalization
of the {\hub} by adding phonons. It turns out that
the symmetry of the standard Hubbard model
is sometimes a special case of a more general quantum group symmetry.
More precisely, while the ``magnetic'' $\mbox{SU}(2)$-symmetry is
left unchanged, Montorsi and Rasetti
claimed that the generators of a ``superconductive'' $\mbox{SU}_q(2)$
quantum group commute with
their extended Hamiltonian. We were able to verify this symmetry for an
extended Hubbard model on a
one-dimensional lattice\footnote{ The Hamiltonian in \cite{rasetti} is
given {\it explicitly}\/ only in the one-dimensional case.},
while we found unsurmountable obstructions in the higher dimensional case.
As we will show this is essentially due to ordering problems.
Our task in this article is twofold: we will address quantum symmetries in
general and we will carefully re-examine the Hubbard
model with phonons,
deriving each term on physical grounds to obtain the correct Hamiltonian.

\newsection{Quantum symmetries in quantum mechanics}
\label{quantum}

The role of symmetries in quantum mechanics cannot be underestimated.
Some models (harmonic oscillator, hydrogen atom) were in fact first
solved relying only on symmetries. Symmetries, especially infinite
dimensional ones, serve to provide the constants of motion
that are central to integrable models.

It is interesting to see what happens when the usual notion of
symmetry is relaxed and transformations given by a Hopf algebra
(Quantum Group) are considered.

To simplify the discussion we will use a formalism that avoids
direct reference to Hopf algebraic methods.
As given data we take a $*$-Hopf algebra $\U$,
its dual Hopf algebra $\U^*$ and a $*$-algebra $\A$ generated by
quantum mechanical operators that act on a Hilbert space $\H$.
The generators of quantum symmetry transformations live in $\U$.
Here we typically have a one or more parameter deformation
of the universal enveloping algebra of a Lie algebra in mind.
The elements of the dual Hopf algebra $\U^*$ play the role of functions
on the Quantum Group. The only difference to the classical case is that
these functions no longer commute.

\paragraph{Unitary representation}

The elements of $\U$ should act on $\H$. We need a unitary representation
$\rho$ on $\H$ that realizes $\U$ in the operator algebra $\A$. Such
a representation shall be a linear $*$-preserving map
\beq
\rho\,:\: \U  \rightarrow  \A,\qquad
\rho(x)\,:\: \H  \rightarrow  \H,\qquad
\rho(x)^\dagger  =  \rho(x^*)
\eeq
that is also an algebra homomorphism\footnote{{\em Remark:}\ $\rho(x)\rho(y) =
\rho(z) \Leftrightarrow \rho(z) = \rho(x y) \Leftarrow z = x y$, but {\em
not}: ``$\Rightarrow z = x y$''}
\beq
\rho(x y) = \rho(x) \rho(y).
\eeq
Here is an example:

\paragraph{Magnetic and Superconductive SU${}_{(q)}(2)$}
The algebra of SU${}_q$(2) is generated by
$X^+$, $X^- = (X^+)^*$ and $H = H^*$ with deformed
commutation relations
\beq
[H,X^{\pm}]=\pm 2 X^{\pm}, \qquad [X^+,X^-]=\frac{q^{H}-q^{-H}}{q- q^{-1}};
\qquad q \in \mbox{\bf R}\backslash\{0\} .
\label{uqsu2}
\eeq
As can be checked by direct computation this algebra has the same
representation by $2 \times 2$ matrices as the undeformed SU$(2)$,
namely
\beq
X^+ \mapsto \left(\begin{array}{cc}0&1\\0&0\end{array}\right),\qquad
X^- \mapsto \left(\begin{array}{cc}0&0\\1&0\end{array}\right),\qquad
H \mapsto \left(\begin{array}{cc}1&0\\0&-1\end{array}\right).
\label{nuqsu2}
\eeq
{}From this matrix representation we can find a unitary representation
of the algebra (\ref{uqsu2}) by creation and
annihilation operators using the following simple observation:
\begin{quote} Let $c_i^\dagger$, $c_i$ be fermionic or bosonic
creation and annihilation
operators and $m_{ij}$, $n_{ij}$ numerical matrices
with the same (finite) index set as the $c^\dagger$, $c$,
then $[c^\dagger\cdot m \cdot c , c^\dagger\cdot n \cdot c ] =
c^\dagger \cdot [ m \stackrel{\cdot}{,} n ] \cdot c$.\end{quote}
If we take for instance $c^\dagger_i \in \{ b^\dagger_\uparrow ,
b^\dagger_\downarrow \}$ and $c_i \in \{ b_\uparrow ,
b_\downarrow \}$ and the matrices from (\ref{nuqsu2}) we find
the generators of the ``magnetic'' SU(2):
\beq
\rho_m(X^+) = \bu^\dagger \bd,\quad \rho_m(X^-) = \bd^\dagger \bu,\quad
\rho_m(H) = (\bu^\dagger \bu - \bd^\dagger \bd).
\label{muqsu2}
\eeq
Switching $\bd^\dagger \leftrightarrow \bd$ does not change the
algebra of the $c^\dagger$, $c$ (the $b^\dagger$, $b$ are fermionic!)
but gives another unitary representation---the ``superconductive''
SU${}_{(q)}(2)$:
\beq
\rho_s(X^+) = \bu^\dagger \bd^\dagger,\quad \rho_s(X^-) = \bd \bu,\quad
\rho_s(H) = (\bu^\dagger \bu + \bd^\dagger \bd - 1).
\label{suqsu2}
\eeq
(These expressions hold also for $q \neq 1$ because $[\rho_{m/s}(H)]^3 =
\rho_{m/s}(H)$.)
These  generators implement both (local) SU$(2)$ and (local) SU${}_{q}(2)$
for a {\em single lattice site}. When we deal with generators
that act on the {\em whole lattice}\/ the ``$q$'' reappears and consequently
(global) SU$(2)$ and (global) SU${}_{q}(2)$ do no longer coincide.

\noindent {\bf Note:} In the sequel we will not write the
symbol ``$\rho$'' explicitly; its presence is implied by context.

\paragraph{Transformation of States and Operators}

The key to a simple description of quantum symmetries is the canonical
element of $\U \ot \U^*$ sometimes also called the ``universal T''
\cite{FRT}
\beq
\T \equiv \sum_i e_i \ot f^i \in \U \ot \U^*;
\eeq
here $e_i$ and $f^i$ are (formal) dual linear bases of $\U$ and $\U^*$
respectively.
Everything else we need to know about $\T$ here is that it is invertible
and unitary:
\beq
\T^* \equiv \sum_i e_i^* \ot f^i{}^* = \T^{-1}.
\eeq
States $|\psi\ra \in \H$ corresponding to a single site\footnote{%
Statements for ``single sites'' and ``multiple sites''
of a lattice obviously
apply also to a broader context of tensor products of states---for
instance to single/multi-particle states.}
transform via multiplication by $\T$:
\beq
|\psi\ra \mapsto \T |\psi\ra .
\eeq
Operators $\Op \in \A$ consequently transform by conjugation
\beq
\Op \mapsto \T \Op \T^{-1} = \T \Op \T^*.
\eeq
States and operators can have full quantum symmetries,
{\em i.e.}\/ they can be invariant under all of $\U$. This
is the case if respectively:
\beq
\begin{array}{rcl}
\T |\psi\ra & = & 1 \cdot |\psi\ra \\
\T \Op \T^{-1} & = & \Op  \cdot 1
\end{array}\qquad\mbox{(conditions
for {\em full}\/ symmetry)}
\eeq

When we deal with a lattice, there
is a $\T_i$ for each of its sites~$i$.
Transformations of several sites (the whole lattice)
{\em i.e.}\/ of states $|\psi^{(N)}\ra \in \H^{\ot N}$
and operators $\Op^{(N)} \in \A^{\ot N}$
are also possible. These are performed with products
(in the function part) of the $\T_i$,
\beq
\T^{(N)} = \T_1 \T_2 \ldots \T_N \equiv
\sum_{i_1,i_2,\ldots,i_N} e_{i_1} \ot e_{i_2} \ot\ldots\ot e_{i_N} \ot
f^{i_1} f^{i_2} \ldots f^{i_N},
\eeq
so that
\beq
|\psi^{(N)}\ra \mapsto \T^{(N)} |\psi^{(N)}\ra, \qquad
\Op^{(N)} \mapsto \T^{(N)} \Op^{(N)} (\T^{(N)})^{-1},
\eeq
with $(\T^{(N)})^{-1} = \T_N^{-1} \T_{N-1}^{-1} \ldots \T_1^{-1}$.
Note that the order of the $\T_i$ in $\T^{(N)}$ is important because
the $f^i$ (in the function part of $\T$) are not commutative by assumption
for a quantum group.

\paragraph{Full quantum symmetry}
In the following sections we will be interested in quantum symmetries of the
Hamiltonian. A Hamiltonian $h \in \A$ has a full ``local''
symmetry under $\U$ (at site $i$) if
\beq
\T_i h \T_i^{-1} = h \cdot 1; \label{localU}
\eeq
it consequently has a full ``global''  symmetry under $\U$ (on the whole
lattice)
if
\beq
\T^{(N)} h \Big(\T^{(N)}\Big)^{-1} = \T_1 \T_2 \ldots \T_N h
\T_N^{-1} \T_{N-1}^{-1} \ldots \T_1^{-1} = h \cdot 1. \label{globalU}
\eeq
In this formalism it is very easy to see that
both conditions can also be expressed in terms of commutators,
namely
\beq
\Big[ \T_i , h \Big] = 0 \qquad \mbox{and} \qquad
\Big[\T^{(N)} , h \Big] = 0
\eeq
respectively.

\paragraph{Specified transformations}
Often it is important to describe transformations given
by specific elements of the Hopf algebra $\U$. So far the transformations
were unspecific; their result still contained a part in $\U^*$ {\em i.e.}\/
a ``function on the quantum group''
\beq
\mbox{\it e.g.} \quad \T |\psi\ra
\equiv \sum_i \rho(e_i)|\psi\ra \ot f^i  \in \H \ot \U^* \quad
\mbox{and similar} \quad
\T \Op \T^{-1} \in \A \ot \U^*.
\eeq
A transformation specified by an element $\kappa\in \U$ is
obtained by evaluating these function parts on $\kappa$;
this operation will be denoted by ``$|_\kappa$''.
(You may think of it as ``plugging-in'' of the
transformation parameters.)
The action (denoted by ``$\tr$'') of $\kappa$ on a
state $|\psi\ra$ is then given by
\beq
\kappa \tr |\psi\ra = \T|_\kappa\:|\psi\ra = \kappa|\psi\ra \equiv
\rho(\kappa)|\psi\ra \label{217}
\eeq
simply because $\T$---being the canonical element---satisfies $\T|_\kappa
= \sum_i e_i \cdot f^i(\kappa) = \kappa$ by definition.
Similarly
\bea
\kappa \tr \Op & = & \T \Op \T^{-1} \Big|_\kappa ,\\
\kappa \tr |\psi^{(N)}\ra & = & \T^{(N)}\Big|_\kappa |\psi^{(N)}\ra ,
\label{219}\\
\kappa \tr \Op^{(N)} & = & \T^{(N)} \Op^{(N)} (\T^{(N)})^{-1} \Big|_\kappa .
\label{220}
\eea

The result of contracting the function part of $\T^{(N)} =
\T_1 \T_2 \ldots \T_N$ with $\kappa$ gives a prescription
(denoted by $\Delta^{(N-1)}(\kappa)$
and called the $N-1$-fold coproduct
\footnote{The coproduct
$\Delta$ did not enter the
formalism
as additional input here; it rather follows from Hopf algebra axioms that
\begin{eqnarray*}
\T_1 \T_2 \Big|_\kappa & \equiv & \sum_{i,j} e_i \ot e_j \ot (f^i f^j)(\kappa)
\; = \; \sum_k \Delta(e_k) \ot f^k(\kappa) \; = \; \Delta(\kappa),\\
\T_1 \T_2 \T_3 \Big|_\kappa & = & (\Delta \ot i\!d)\Delta(\kappa) =
(i\!d \ot \Delta)\Delta(\kappa) =: \Delta^{(2)}(\kappa),\\
& \vdots &\\
\T_1 \T_2 \ldots \T_N \Big|_\kappa & = &  \Delta^{(N-1)}(\kappa).
\end{eqnarray*}
The coproducts of a given Hopf algebra are part of the defining relations.
Here are the coproducts for the generators of the algebra
(\ref{uqsu2}):
$$
\Delta(H) = H \ot 1 + 1 \ot H, \quad
\Delta(X^\pm) = X^\pm \ot q^{-H/2} + q^{H/2} \ot X^\pm.
$$
Coproducts of other elements can be computed from this using the fact
that $\Delta$ is an algebra homomorphism.
The other objects that constitute a Hopf algebra are the
antipode $S$ and the counit $\epsilon$. They enter our formalism
via $\T^{-1}|_\kappa = S(\kappa)$ and $1|_\kappa = \epsilon(\kappa)$.
Note that $\epsilon(\kappa)$ is a number.\\
Let $\Delta(\kappa) \equiv \kappa_{(1)} \ot \kappa_{(2)}$; then
$\T \Op \T^{-1}|_\kappa = \sum_{i,j} \rho(e_i) \Op \rho(S e_j) \ot (f^i
f^j)(\kappa) = \rho(\kappa_{(1)}) \Op \rho(S \kappa_{(2)})$.
This action and Hopf-expressions corresponding to equations
(\ref{219}--\ref{220}) are {\em e.g.}\/ discussed in~\cite{identical}.}
)
how to distribute $\kappa$ over several tensor factors:
\beq
\T^{(N)}\Big|_\kappa = \Delta^{(N-1)}(\kappa) \in \U^{\ot N}.
\label{delta}
\eeq
It is clear that there cannot be one simple rule for all
of $\U$---not even in the classical case;
$\Delta(\kappa) = \kappa\ot 1 + 1 \ot \kappa$ for
instance holds only for ``infinitesimal'' $\kappa$.
The added difference of the quantum case is that then $\Delta(\kappa)$
will in general be not symmetric.

\paragraph{Partial quantum symmetry}

The full quantum symmetries (\ref{localU}) and (\ref{globalU}) are
equivalent to
\beq
\T_i h \T_i^{-1}\Big|_\kappa =
h \cdot 1_\kappa \ \forall \kappa \in \U \quad
\mbox{and} \quad \T^{(N)} h (\T^{(N)})^{-1} \Big|_\kappa =
h \cdot 1_\kappa  \ \forall \kappa \in \U
\eeq
respectively. We have seen that these full symmetries could
be expressed in terms of commutators.
As a further illustration of the formalism  we will
briefly study the case where $\kappa$ does not
range over all of $\U$ but only over a subset $\P \subset \U$.
The question is: when is
\beq
\T h \T^{-1} \Big|_\kappa = h \cdot 1 \Big|_\kappa \quad \forall \kappa \in \P
\qquad \mbox{({\em partial} quantum symmetry)}
\eeq
equivalent to
\beq
\Big[ \T , h \Big] \Big|_\kappa = 0 \quad \forall \kappa \in \P
\eeq
for an arbitrary Hamiltonian $h$?
A sufficient condition is easily seen to be
\beq
A \,\T \Big|_\kappa = 0 \Leftrightarrow A \cdot 1 \Big|_\kappa = 0
\quad \forall \kappa \in \P
\eeq
for all operators $A$ ($\in \A \ot \U^*$).
This can be translated into a condition on the coproducts
of elements in $\P$:
\beq
\Delta(\P) \subset \P \ot \U.
\eeq

\newsection{A generalized Hubbard model}
\label{sectHUB}

Following \cite{rasetti} we will retain the local electron-term (\ref{elloc}),
and add to it the standard hamiltonian for
the phonons and a phonon-electron interaction term
\beq
\hh=H_{el}^{(loc)}+H_{ph}+H_{el-ph}.
\eeq
We suppose that the phonons are described by a set of decoupled
Einstein oscillators with the same frequency $\omega$
\beq
H_{ph}=\sum_i \left(\frac{\vec p_i{}^2}{2M}
+\frac{1}{2} M\omega^2 \vec y_i{}^2\right)
\eeq
where $\vec p_i$ and $\vec y_i$ obey canonical commutation
relations as usual.
The expression for the phonon-electron term  is the one
given by Hubbard \cite{hubbard}
\beq
H_{el-ph}=\sum_{ij} \sum_{\sigma} \int d^Dr\: \Psi^* (\vec r-\vec R_i)
\left(-\frac{\hbar^2 \nabla^2}{2m}+V(\vec r,\{\vec R_l\})\right)
\Psi (\vec r-\vec R_j)
\: b^{\dagger}_{j \sigma} b_{i \sigma}
\label{helph}
\eeq
where $\Psi(\vec r-\vec R_i)$ is the Wannier electron wave function
centered around the ion at
$\vec R_i$, while $b^{\dagger}_{j \sigma}$, $b_{i \sigma}$ are
fermionic creation
and annihilation operators.
(In this context the Wannier functions will be approximated by atomic
orbitals.)
To take account of the ion oscillations around their equilibrium positions,
the arguments of the Wannier functions and of the potential $V$
in the integral must be shifted:
\[ \vec R_k \rightarrow \vec R_k+\vec y_k, \qquad (k = i,j,l).
\]
The term obtained from
the potential $V$ in (\ref{helph}) has significant contribution only
for $i=j=l$  ({\em i.e.}\/ neglecting all $\vec R_l$ with $l \neq i$ in $V$)
and results by linear variation in a local
electron-phonon interaction term \cite{Hol,FW}
\beq
H_{el-ph}^{(loc)}=-\vec{\lambda} \cdot \sum_i (\niu + \nid) \vec{y_i}
\eeq
and a term that contributes to $\mu$ in (\ref{elloc}).
The non-local electron-phonon interaction term is crucial
in the approach of Montorsi and Rasetti. We would like to give a
derivation leading directly
to the exponential form, which is necessary for
the quantum symmetry. (See \cite{BLF} for the derivation of a linear
approximation.)

We shall retain only the nearest
neighbour terms $\langle ij\rangle$ in the kinetic energy term of
(\ref{helph});
this assumes negligible overlap between
all other atomic orbitals:
\beq
H_{el-ph}^{(non-loc)}=\sum_{\la i,j \ra}
T_{ij} b^{\dagger}_{j \sigma} b_{i \sigma}
\eeq
with $T_{ij} = T^\dagger_{ji}$ given by
\beq
T_{ij}=\int d^Dr \:\Psi^* (\vec r-\vec R_i-\vec y_i) \left(-\frac{\hbar^2
\nabla^2}{2m}\right)  \Psi (\vec r-\vec R_j-\vec y_j).
\eeq
Assuming that $\Psi$ has finite support, it is possible
to shift the integration variable
\[ \vec r \rightarrow \vec r-\vec R_j-\vec y_j.
\]
With this substitution $T_{ij}$ becomes a function only of
$\vec a_{ij} \equiv (\vec R_i + \vec y_i) - (\vec R_j + \vec y_j)$:
\beq
T_{ij}=\int d^Dr \:\Psi^* (\vec r- \vec a_{ij})
\left(-\frac{\hbar^2 \nabla^2}{2m}
\right)
\Psi (\vec r)=T(\vec a_{ij}).
\eeq
The atomic orbitals show an asymptotic exponential decay
\beq
\Psi(\vec r) \sim e^{-\zeta | \vec r | }
\eeq
and we have hence (approximatly)
\beq
\nabla_{\vec a_{ij}} T(\vec a_{ij})= \int d^Dr \:\zeta
\frac{(\vec r-\vec a_{ij})}{| \vec r-\vec a_{ij} |}\:
\Psi^* (\vec r- \vec a_{ij}) \:\frac{\hbar^2 \nabla^2}{2m}\:\Psi(\vec r).
\eeq
Again, due to the rapid exponential decay of $\Psi(\vec r)$,
we can neglect $\vec r$ in $|\vec r - \vec a_{ij}|$
so that
\beq
\nabla_{\vec a_{ij}} T(\vec a_{ij})=
-\zeta \frac{\vec a_{ij}}{|\vec a_{ij}|} T(\vec a_{ij})
\eeq
which integrates to:
\beq
T(\vec a_{ij})=T_0 e^{-\zeta | \vec a_{ij} |}.
\eeq
$|\vec a_{ij}|=|\vec R_i-\vec R_j+\vec y_i-\vec y_j|$ can be expanded using
$|\vec y_i-\vec y_j| \ll |\vec R_i-\vec R_j|$
such that finally
\beq
T_{ij}=t \,\exp\left(-\zeta \frac{(\vec R_i-\vec R_j)}{|\vec R_i-\vec R_j|}
(\vec y_i-\vec y_j)\right)
\eeq
with a new constant $t = T_0 \exp(-\zeta|\vec R_i - \vec R_j|)$.
Note that the term
$$\vec R_{ij} \equiv -\frac{(\vec R_i-\vec R_j)}{| \vec R_i-\vec R_j |}$$
always has the
same module and that in the one-dimensional case it just amounts to a sign.
$| \vec R_i- \vec R_j |$ is the interatomic distance at equilibrium so that
it does not depend on $i,j$.
The complete non-local electron-phonon interaction
term in the Hamiltonian is
\beq
H_{el-ph}^{(non-loc)}=t \sum_{\la i,j\ra} \sum_{\sigma} \exp\{\zeta \vec R_{ij}
\cdot (\vec y_i- \vec y_j)\}
\, \, \bbjs \bis \label{elphnonloc}
\eeq
and the full Hamiltonian of the Hubbard model with phonons is
\begin{eqnarray}
\lefteqn{H_{\mbox{\scriptsize Hub}}  =}\nonumber\\
&& u \sum_i \niu \nid-\mu \sum_{i,\sigma} \nis
+\sum_i \left(\frac{\vec p_i^2}{2M} +\frac{1}{2} M\omega^2
\vec y_i^2\right)
-\vec{\lambda} \cdot \sum_i (\niu + \nid) \vec{y_i}\nonumber\\
&&+ \, \left(t \sum_{\langle i<j\rangle} \sum_{\sigma} \exp\{\zeta \vec R_{ij}
\cdot (\vec y_i- \vec y_j)\}
\:\bbjs \bis \, + \, h.c. \right) .\label{fulhub}
\end{eqnarray}

The Hamiltonian considered in \cite{rasetti} is formally obtained (in the
one-dimensional case---see remark below) from (\ref{fulhub})
by a similarity transformation (half of a Lang-Firsov transformation
\cite{LF})
 on the fermionic operators
$\bbjs$ and $\bis$ only:
\beq
a_{i \sigma} \equiv U(\vec{\kappa})     b_{i \sigma} U^{-1}(\vec\kappa)
\label{314}
\eeq
\beq
a_{j \sigma}^{\dagger} \equiv U(\vec\kappa)
b_{j \sigma}^{\dagger} U^{-1}(\vec\kappa)
\label{315}
\eeq
with a unitary operator
\beq
U(\vec\beta) \equiv \exp\left(i \vec\beta \cdot \sum_{l,\sigma} \vec p_l \,
n_{l \sigma}\right)
\label{transf}
\eeq
that depends on a set of constant parameters $\beta^k,\ k = 1,\ldots,D$.
While this transformation does not change the number operators
$\nlu$ and $\nld$,
it results in an exponential factor in $\vec p_i-\vec p_j$ for
\beq
b_{j \sigma}^{\dagger} b_{i \sigma}=e^{i \vec\kappa
\cdot (\vec p_i-\vec p_j)}   a_{j \sigma}^{\dagger} a_{i \sigma},
\eeq
so that
the hopping term is now given by
\beq
H_{el-ph}^{(non-loc)}=t \sum_{\la i<j \ra} \sum_{\sigma}
\exp\{\zeta \vec R_{ij}
\cdot (\vec y_i- \vec y_j)\}
\exp\{i\vec{\kappa}\cdot (\vec p_i-\vec p_j)\}\, \, \aajs \ais + h.c.
\eeq
or, combining the exponentials,
\beq
H_{el-ph}^{(non-loc)}=t  \sum_{\la i<j \ra} \sum_{\sigma} e^{-\hbar
\zeta \vec R_{ij} \cdot \vec \kappa} \exp\{\zeta \vec R_{ij}
\cdot (\vec y_i-\vec y_j)+i\vec{\kappa} \cdot(\vec p_i-\vec p_j)\}
\, \, \aajs \ais + h.c.
\eeq
\paragraph{Remark:} Note that while the $\vec y_i$ commute with
$\bbis$, $\bis$, they do not commute with the new fermionic
creation and annihilation operators $\aais$, $\ais$ as defined in (\ref{314},
\ref{315}). The authors of \cite{rasetti}, however, assumed commutativity
between
the fermionic operators and coordinates of the ions. In order to be able to
connect to their work we will formally replace the $\vec y_i$ in
$H_{\mbox{\scriptsize Hub}}$ with new coordinates $\vec x_i$ that do commute
with the $\aais$, $\ais$.
(The $\vec x_i$  will hence no longer commute with the $\bbis$, $\bis$.)
This will of course modify the hamiltonian. The hamiltonian that we {\em
will}\/ work with in the next section is:
\beq
H = H^{(loc)} + H^{(non-loc)},\label{rmham1}
\eeq
with
\beq
H^{(loc)} =
u \sum_i \niu \nid-\mu \sum_{i,\sigma} \nis
+\sum_i \left(\frac{\vec p_i^2}{2M} +\frac{1}{2} M\omega^2
\vec x_i^2\right)
-\vec{\lambda} \cdot \sum_i (\niu + \nid) \vec{x_i},\label{rmham2}
\eeq
\beq
H^{(non-loc)} = t \sum_{\la i<j \ra} \sum_{\sigma}
\exp\{\zeta \vec R_{ij}
\cdot (\vec x_i- \vec x_j)\}
\exp\{i\vec{\kappa}\cdot (\vec p_i-\vec p_j)\}\, \, \aajs \ais + h.c.
\label{hnonloc}
\eeq
The relation of this Hamiltonian with the one of the Hubbard model with phonons
(\ref{fulhub}) will be discussed in section~\ref{discussion}.
The fact that $H_{\mbox{\scriptsize Hub}}$ and $H$ are inequivalent can for
instance be seen by noting that the expression
$$
\tilde T_{ij}=t \exp\{\zeta \vec R_{ij} \cdot (\vec x_i- \vec x_j)\}
\exp\{i\vec{\kappa}\cdot (\vec p_i-\vec p_j)\}
$$
for the hopping amplitude in (\ref{hnonloc}) does not satisfy
the condition $\tilde T_{ji}=\tilde T_{ij}^{\dagger}$
so that $\sum_{\la i,j \ra} \tilde T_{ij} \aajs \ais$ is no longer hermitean.

\newsection{Superconductive U${}_q\,$su(2)}
\label{symmetry}

The local superconductive $U_q(su(2))$ is given by
\bea
\rho_s(X^+) = \klu = b_{l \uparrow}^\dagger b_{l \downarrow}^\dagger
=e^{-i \vec{\Phi} \cdot \vec p_l} \aalu \aald
\label{klu} \\
\rho_s(X^-) = \kld= b_{l \downarrow} b_{l \uparrow}
=e^{i \vec{\Phi} \vec p_l}\ald \alu =(\klu)^{\dagger}
\label{kld} \\
\rho_s(H) = 2 \klz =\nlu + \nld -1
\label{klz}
\eea
These are the generators for transformations of an individual lattice site~$l$,
as defined in (\ref{suqsu2}). They are expressed in terms of the
operators $b_{l \sigma},b_{l \sigma}^{\dagger}$. In order to compute
the commutation relations
with hamiltonian
(\ref{rmham1}) it is necessary to express
them in terms of operators $a_{l\sigma},a_{l\sigma}^{\dagger}$ as introduced
in (\ref{314}) and (\ref{315}). The parameter $\vec{\beta}$ appearing in
(\ref{transf}),
on which the transformation depends, is chosen here to be $\vec{\Phi}/2$ and
for the moment it should be regarded as a free parameter
which will be determined by the commutation relations.
We will see later (\ref{kappa}) that consistently with the choice
$\vec{\beta}=\vec{\kappa}$ made in
equations (\ref{314}) and (\ref{315}), the commutation relations will
require $\vec{\Phi}=2 \vec{\kappa}$.
(Notice that
$n_{l \sigma}=a_{l\sigma}^{\dagger} a_{l\sigma}=
b_{l \sigma}^{\dagger} b_{l\sigma}.$)

To describe the symmetries of the Hubbard model with phonons it is
necessary to consider two distinct representations of the superconductive
$U_q(su(2))$ for different lattice sites. One ($\rho_s^+$) is equal to
$\rho_s$, the other ($\rho_s^-$) differs from $\rho_s$ by a minus sign on the
generators $X^\pm$:
\beq
\rho^{\pm}_s(X^{+})=\pm \rho_s(X^{+}), \quad
\rho^{\pm}_s(X^{-})=\pm \rho_s(X^{-}), \quad
\rho^{\pm}_s(H)=\rho_s(H).
\eeq
For each lattice site $l$ a sign $\sigma(l) \in \{ 1,-1 \}$ is chosen
and the representation $\rho^+_s$ or $\rho^-_s$ is associated
to it depending on whether $\sigma(l)=1$ or $\sigma(l)=-1$ respectively.
The local commutation relations are not affected by this choice.
The sign will however be crucial for the global commutation relations.
Hence, for the moment we will not specify a rule for
assigning a representation
to a given site, but we will see later (\ref{signum})
that sites corresponding to nearest
neighbours must have opposite representations $\rho^+$ and $\rho^-$.
This is exactly what happens in the classical case \cite{yangzhang}.
For orthogonal (square) lattices
a choice of the sign which implements this condition
is
\beq
\sigma(l)=(-1)^{\| l \|}
\eeq
where $\| l \|=\sum_{n=1}^D l_n$ is the length of the index $l$
which labels the site $l$.

For the moment we will choose an
arbitrary ordering of the lattice sites.
Choosing an ordering is necessary to be able to define a tensor
product and hence to construct a global symmetry.
According to the definition of the coproduct in $U_q(su(2))$
(see \ref{delta} and footnote 4)
\bea
& &\Delta(X^+) =e^{\frac{1}{2} \alpha H}
\otimes X^+ + X^+ \otimes e^{-\frac{1}{2} \alpha^* H}, \label{delt1}\\
& & \Delta(X^-)=e^{\frac{1}{2} \alpha^* H}
\otimes X^- + X^- \otimes e^{-\frac{1}{2} \alpha H}=(\Delta(X^+))^{\dagger}, \\
& & \Delta(H)=H \otimes 1 + 1 \otimes H, \label{delt3}
\eea
where the deformation parameter is chosen to be
$q=e^{\alpha}$ and $\alpha $
is a complex para\-meter to be determined by the
commutation relations and through the representations $\rho_s^\pm$ we obtain
the generators
of global superconductive
$U_q(su(2))$
\bea
\ku & = & \bigotimes_{l} \rho_s^{\sigma(l)} (\Delta^{(N-1)} (X^+)), \\
\kd & = & \bigotimes_{l}
\rho_s^{\sigma(l)} (\Delta^{(N-1)} (X^-))= (\ku)^{\dagger}, \\
\kz & = & \bigotimes_{l} \rho_s^{\sigma(l)} (\Delta^{(N-1)} (H)),
\eea
where $N$ is the number of lattice sites.
Using (\ref{delt1}--\ref{delt3}) these generators are computed to be
\bea
\ku=\sum_l \sigma(l) \prod_{r<l} \, \, e^{\alpha \krz} \,
\,\klu \, \,
\prod_{r>l} \, \, e^{-\alpha^* \krz},
\label{ku} \\
\kd=\sum_l \sigma(l)  \prod_{r<l} \, \, e^{\alpha^* \krz} \, \,
\kld \, \, \prod_{r>l}
\, \, e^{-\alpha \krz} =(\ku)^{\dagger},
\label{kd} \\
\mbox{and} \qquad \kz=\sum_l \klz.
\label{kz}
\eea

\paragraph{Local commutation relations}
The local part of the Hamiltonian commutes with the local generators
\beq
\left[ \klu,H^{(loc)}  \right]= \left[ \kld,H^{(loc)} \right]=
\left[ \klz,H^{(loc)} \right]=0
\eeq
if the following conditions hold
\bea
\vec\Phi &=&\frac{2\vec\lambda}{\hbar M \omega^2},\label{loccond1}\\
\mu &=&\frac{u}{2} - \frac{1}{4} M \omega^2 \hbar^2 \Phi^2 \: = \:
\frac{u}{2} -\frac{\vec{\lambda}^2}{M\omega^2}. \label{loccond2}
\eea

\paragraph{Global commutation relations}

The fact that $\kz$ commutes with $H^{(non-loc)}$ given by
(\ref{hnonloc}) is immediate.
We must calculate
\bea
\lefteqn{\left[ \ku,H^{(non-loc)} \right]=}
\label{comm0} \\
& \left[
{\displaystyle \sum_{l} \sigma(l) \prod_{r<l}e^{\alpha \krz}
\kju \prod_{r>l}e^{-\alpha^* \krz},
 {\displaystyle t \sum_{\la i < j \ra} \sum_{\sigma}}}
 e^{\zeta \vec R_{ij} \cdot (\vec x_i- \vec x_j)} \, \, e^{i \vec{\kappa}
\cdot (\vec p_i- \vec p_j)}
\, \, \aajs \ais  \right] .\nonumber
\eea
It can be seen that
\bea
&&\left[ e^{-i \vec{\Phi} \cdot \vec p_l},
e^{\zeta \vec R_{ij} \cdot (\vec x_i-\vec x_j)} \right] =
2 \sinh({1 \over 2}\zeta\hbar \vec R_{ij}\cdot \vec\Phi)
(\delta_{l,j}-\delta_{l,i}) e^{-i \vec{\Phi} \cdot \vec p_l +
\zeta \vec R_{ij}\cdot(\vec x_i-\vec x_j)}, \label{comm3}\\
&&\left[\aalu \aald,\aaju \aiu + \aajd \aid\right] =
-\delta_{l,i} (\aaju \aaid + \aaiu \aajd),
\label{comm4}\\
&&e^{\alpha \kiz} = 1+2 \kiz (1-e^{-\frac{\alpha}{2}})+2 \niu \nid
\left( \cosh({\alpha \over 2})-1
\right)
\label{comm5}
\eea
and, using equation (\ref{comm5}),
\bea
\lefteqn{\left[ e^{\alpha \klz},\aaju \aiu +\aajd \aid \right]}
\nonumber \\
& = & (\aaju \aiu +\aajd \aid) \big( \delta_{l,j}-\delta_{l,i} \big)
(1- e^{-\frac{\alpha}{2}}) \label{comm6}\\
& + & \left( \delta_{l,j} (\aaju \aiu \njd +\nju \aajd \aid)
-\delta_{l,i} (\aaju \aiu \nid +\niu \aajd \aid) \right)
(e^{\frac{\alpha}{2}} + e^{-\frac{\alpha}{2}} -2).
\nonumber
\eea
We introduce the abbreviation
\beq
Z_{ij}=\sigma(i) \ei \prod_{r<i, r \not= j}e^{\alpha \krz}
\prod_{r>i, r \not= j}e^{-\alpha^* \krz}.
\eeq
Splitting  the commutators, evaluating the
expressions that are obtained by the use of
(\ref{comm3}--\ref{comm6}),
and using the delta-functions which appear in (\ref{comm3}),
(\ref{comm4}) and (\ref{comm6}) to perform some of the sums,
it can be seen that (\ref{comm0}) becomes
\bea
\lefteqn{\left[ \ku,H^{(non-loc)} \right]=}
\nonumber \\
& & t {\displaystyle \sum_{\la i<j \ra}} e^{-\hbar
\zeta \vec R_{ij} \cdot \vec \kappa}
\Bigg\{
(\aaid \aaiu \aajd \aaju  \aiu \ajd
-\aaid \aaiu \aajd \aaju \aid \aju) \nonumber \\
& \times &  \Bigg[ Z_{ij} \left( 2 \cosh \left( \frac{1}{2} \vec R_{ij} \cdot
\vec{\Phi} \zeta \hbar \right)
-  2 \cosh \left(\frac{1}{2}   \vec R_{ij} \cdot
\vec{\Phi} \zeta \hbar + \frac{1}{2} \alpha^* \right) \right) \nonumber \\
& + & Z_{ji} \left(  2 \cosh \left( \frac{1}{2} \vec R_{ij} \cdot
\vec{\Phi} \zeta \hbar \right)
-  2 \cosh \left(\frac{1}{2}   \vec R_{ij} \cdot
\vec{\Phi} \zeta \hbar + \frac{1}{2} \alpha \right)
\right) \Bigg] \nonumber \\
& + &  (\aaiu \aajd \aaju \aju + \aaid \aajd \aaju \ajd)
\epi \Bigg[ Z_{ij} (\eajim -1)  + Z_{ji}
\left( e^{-\vec R_{ij} \cdot
\vec{\Phi} \zeta \hbar} \eaijm  -1 \right) \Bigg] \nonumber \\
& + & (\aaid \aaiu \aajd \aid + \aaid \aaiu \aaju \aiu) \epj \Bigg[  Z_{ij}
(e^{\vec R_{ij} \cdot  \vec{\Phi} \zeta \hbar}\eajim - 1)
+ Z_{ji} (\eaijm-1) \Bigg] \nonumber \\
& + & (\aaid \aaju -\aaiu \aajd) \Bigg[ Z_{ij} \epi \eajim
+ Z_{ji} \epj \eaijm ) \Bigg] \Bigg\} \nonumber \\
& + & {\displaystyle \sum_l \sum_{\la i,j \ra ,i<l<j}}
\sigma (l) e^{-\hbar \zeta \vec R_{ij}
\cdot \vec \kappa} e^{-i \vec{\Phi} \cdot \vec p_l} \aalu \aald
{\displaystyle
\prod_{r<l,r \not=i} e^{\alpha \krz}}
{\displaystyle \prod_{r>l,r \not=j}} e^{-\alpha^* \krz} \nonumber \\
& \times & \left[ e^{\alpha \kiz} e^{-\alpha^* \kjz},
e^{i \vec{\kappa} \cdot  (\vec p_i-\vec p_j)
+\zeta \vec R_{ij} \cdot (\vec x_i-\vec x_j)} \aaju \aiu
+\aajd \aid +h.c. \right].
\eea
There are 2 sums containing 6 fermionic operators, 4 sums containing
4 fermionic operators, and 2 sums containig 2 fermionic operators.
These  sums must all be separately 0, because they depend on
different numbers of such operators and hence are linearly
independent.
Let's study the term containing $\aaid \aaju -\aaiu \aajd$:
\beq
{\displaystyle \sum_{\la i<j \ra}}
(\aaid \aaju -\aaiu \aajd)  \nonumber
\Bigg[ Z_{ij} \epi \eajim + Z_{ji} \epj \eaijm ) \Bigg]
\label{termaa}
\eeq
The above sum can vanish only if each term with fixed $i,j$
is separately 0, because there are no other terms which
contain $\aaid \aaju -\aaiu \aajd$.
Therefore it is necessary that the expression between the
square brackets is 0.
For this reason we must require
\beq
Z_{ij} \epi \eajim + Z_{ji} \epj \eaijm =0.
\eeq
This is equivalent to the set of equations
\beq
Z_{ij}=-Z_{ji},
\eeq
\beq
\epi \eajim = \epj \eaijm,
\eeq
which in turn imply---($i,j$ are nearest neighbours)
\beq
\sigma(i)=-\sigma(j),
\label{signum}
\eeq
\beq
\ei = \ej,
\label{eij}
\eeq
\beq
 {\displaystyle \pij=\pji,}
\label{prod}
\eeq
\beq
e^{-\frac{i}{2} Im \alpha}
(\epi e^{\frac{1}{2} Re \alpha} - \epj e^{-\frac{1}{2} Re \alpha})=0.
\label{realfa}
\eeq

Eq. (\ref{signum}) means that nearest neighbours must have opposite
signs. As we have already anticipated, this means that in order for
the global commutation relations to hold, it should be possible
to see the lattice $\Lambda$
on which the model is defined, as the sum of two lattices
$\Lambda_1,\Lambda_2$, such that nearest neighbours are always on
different lattices. This gives a restriction on the possible lattices,
e.g. a triangular lattice could not be chosen.

Eq. (\ref{eij}) implies
$\vec\kappa-\vec\Phi=-\vec\kappa$ and hence
\beq
2 \vec\kappa=\vec\Phi.
\label{kappa}
\eeq
This is one condition that must be satisfied for expression
(\ref{termaa}) to vanish.
In particular it fixes the parameter of the transformation
(\ref{transf}). It turns out, that the parameter has to be the same
as the one used to transform the fermionic operators in the
hamiltonian.

Eq. (\ref{realfa}) implies
\beq
Re \alpha =-\vec R_{ij}\cdot \vec\Phi \zeta \hbar.
\label{alfa}
\eeq
This is the second condition which must be satisfied for expression
(\ref{termaa}) to vanish. It is important to notice that
it is possible to fulfill
this relation only if the ordering of the lattice sites is chosen to be the
lexicographic one. So this imposes a first restriction on the ordering of
the sites.

The strongest relation is (\ref{prod})---it depends crucially on the ordering
chosen for the lattice sites.
In order for (\ref{prod}) to hold it is necessary that
\beq
\prod_{i<r<j}e^{\alpha \krz}=
\prod_{i<r<j}e^{-\alpha^* \krz}.
\label{pro1}
\eeq
Let's apply (\ref{comm5}) to expand the exponentials.
Then we obtain an expression of the type
\beq
1+ 2 (1-e^{\frac{-\alpha}{2}}) \sum_{i<r<j} \krz + \ldots=
1+ 2 (1-e^{\frac{\alpha^*}{2}}) \sum_{i<r<j} \krz + \ldots
\label{pro2}
\eeq
(Here the terms which are indicated with ``$\ldots$'' are at least quadratic
in the $\krz$ and therefore are independent from the
first-order terms which have been written.)
Eq. (\ref{pro2}) shows that in order for relation (\ref{pro1}) to hold,
it is necessary that
$$
e^{-\frac{\alpha}{2}}=e^{\frac{\alpha^*}{2}}\quad \Rightarrow \quad
Re(\alpha)=0.
$$
But this would  mean that the coproduct should be symmetric, and
this is against the hypothesis that there is a true quantum symmetry.

This shows that we must look for a condition on the ordering of the
lattice sites, so that we do not need to require (\ref{pro1}):
There cannot be any site $r$ which satisfies
the condition $i<r<j$ for any couple of nearest neighbours $i,j$.
In other words it is necessary that if $i,j$ is a couple
of nearest neighbours then
\beq
(i < r \Rightarrow j \le r , \quad
i > r \Rightarrow j \ge r) \quad \forall r . \label{cond}
\eeq
However, condition (\ref{cond}) implies that the lattice
$\Lambda$ on which the {\hub} is defined is one-dimensional,
and that the `normal' ordering of the sites is chosen, in which
the sites are numbered from left to right in increasing or
decreasing order.

It can be verified immediately that  condition (\ref{cond})
is {\em sufficient}\/ to guarantee that the sum with
``$ {\sum_l \sum_{\la i,j \ra,i<l<j}}$''
is not present, because there is
no longer any $l$ satisfying
$i<l<j$.
In fact it can be verified with arguments similar to the ones
used to study the term (\ref{termaa}) that (\ref{cond}) is also {\em
necessary}\/
for such a sum to vanish.
Because of (\ref{prod}) and (\ref{eij}) it is possible to combine
the terms which contain the same products of fermionic operators.
Further, it can be seen immediately, that conditions
(\ref{signum}),(\ref{kappa}),(\ref{alfa}) and (\ref{cond}) are also
necessary and sufficient
for the sums with 4 and with 6 fermionic operators to vanish.
Thus in order for the hamiltonian $H^{(non-loc)}$ (\ref{hnonloc}) to
commute with the generators (\ref{ku}--\ref{kz})
of $U_q(su(2))$ the conditions
(\ref{signum}),(\ref{cond}),(\ref{kappa}),(\ref{alfa})
are necessary and sufficient.

\section{Discussion}
\label{discussion}

The computation of the previous section has shown that the Hamiltonian
$H$, given
for a $D$-dimensional lattice in (\ref{rmham1}--\ref{hnonloc}),
commutes with the  generators
of (global) U${}_qsu(2)$, $q \equiv e^{\alpha}$,
provided that local conditions (\ref{loccond1},
\ref{loccond2}), global conditions (\ref{kappa}, \ref{alfa}) and
either $D=1$ or $\alpha =0$ hold. (We will only consider Im$(\alpha) = 0$
here.)
We are now in the position to comment on the symmetries\footnote{
The Hamiltonian will commute with {\em all} elements of U${}_q$SU(2)
provided that its generators do so; this is equivalent to a full quantum
symmetry (under quantum adjoint action),
see section~\ref{quantum}.}
of various
hamiltonians derived from the Hubbard model:

$H_{\mbox{\scriptsize Hub}}$ (\ref{fulhub}) coincides with
(\ref{rmham1}--\ref{hnonloc}) for the particular choice of parameter
$\vec\kappa = 0$. The global conditions
(\ref{kappa}, \ref{alfa}) imply $\vec \Phi =0$,
$\alpha =0$ and hence $q =1$. The local conditions (\ref{loccond1},
\ref{loccond2}) imply $\vec\lambda = 0$ and $\mu = u/2$.
We conclude that the
Hubbard model with phonons (\ref{fulhub})
has no true quantum symmetry---not even in the
one-dimensional case; it has an ordinary SU(2) symmetry provided that there is
no local electron-phonon coupling ($\vec\lambda = 0$).

The Hamiltonian $H$ (\ref{rmham1}) studied in the previous section and
considered by
\cite{rasetti} is formally obtained from the Hubbard model with phonons
by a Lang-Firsov transformation on the fermionic operators only.
(See remark at the end of section~\ref{sectHUB}.)
The essential difference between $H_{\mbox{\scriptsize Hub}}$ and $H$ is that
$H_{\mbox{\scriptsize Hub}}$ uses coordinates $\vec y_i$ that commute with
$\bbis, \bis$ while $H$ is written in terms of new coordinates that commute
with
$\aais$ and $\ais$ (and not with $\bbis, \bis$).
To be able to compare the two Hamiltonians we have to relate the sets of
coordinates. This is simply done by a Lang-Firsov
transformation\footnote{This observation is also
supported by the choice of $K_l^{(\pm)}, K_l^{(z)}$.}
(see \ref{314}--\ref{transf})
\beq
\vec x_i = U(\vec\kappa) \vec y_i U^{-1}(\vec\kappa).
\eeq
The new coordinates are found to be
\beq
\vec x_i  = \vec y_i +
\hbar \vec\kappa \sum_{\sigma} n_{i \sigma},
\eeq
{\em i.e.}\/ the position of the ion at lattice site $i$ is shifted according
to the number of electrons at that site.
Expressing $H$ in terms of $y_i,\bbis, \bis$ we find
\begin{eqnarray}
H_{\mbox{\scriptsize q-sym}} & =
& u' \sum_i \niu \nid-\mu' \sum_{i,\sigma} \nis
+\sum_i \left(\frac{\vec p_i^2}{2M} +\frac{1}{2} M\omega^2
\vec y_i^2\right)\nonumber\\
&&-\,\vec{\lambda}' \cdot \sum_i (\niu + \nid) \vec{y_i}\nonumber
+ \,\Bigg( t \sum_{\langle i<j\rangle} \sum_{\sigma} \exp\{\zeta \vec R_{ij}
\cdot (\vec y_i- \vec y_j)\} e^{-\zeta \hbar\vec R_{ij} \cdot \vec{\kappa}}
\:\bbjs \bis \nonumber\\
&&\times\, \Big(1 + (e^{\zeta \hbar\vec R_{ij}
\cdot \vec\kappa} - 1)n_{i, -\sigma} \Big)
\Big(1 + (e^{-\zeta \hbar\vec R_{ij}
\cdot \vec\kappa} - 1)n_{j, -\sigma} \Big)  \, + \, h.c. \Bigg)
\label{hqsym}
\end{eqnarray}
with a set of new parameters
\begin{eqnarray}
\vec\lambda' & = & \vec\lambda - M \omega^2 \hbar \vec\kappa,\\
u' & = & u -2 \hbar \vec\lambda\cdot\vec\kappa + M \omega^2 \hbar^2 \kappa^2,\\
\mu' & = & \mu + \hbar \vec\lambda\cdot\vec\kappa - 1/2
M \omega^2 \hbar^2 \kappa^2.
\end{eqnarray}
The model given by $H_{\mbox{\scriptsize q-sym}}$ (\ref{hqsym}) has a true
quantum symmetry (in the one-dimen\-sio\-nal case). The local conditions
(\ref{loccond1},  \ref{loccond2}) for quantum symmetry
expressed in terms of  the {\em new} parameters
is
\beq
\vec{\lambda'}=0, \qquad \mu' = u' / 2.
\eeq
There is apparently no local coupling to the phonons and the condition
for symmetry is ``half filling'' as in the standard Hubbard model.

\section*{Acknowledgements}

We thank Julius Wess for suggesting this line of research and
A.~Montorsi and M.~Rasetti for
discussion of their work. We would like to dedicate this article to the memory
of L.~M.~Falicov.

\end{document}